\documentclass[10pt]{article}
\usepackage{epsf,amsfonts,amsthm}
\usepackage{fontenc,indentfirst, delarray,amsfonts,amsmath,amssymb}
\usepackage{graphicx}
\DeclareGraphicsExtensions{ps,eps,ps.gz} \headheight 0.6cm
\makeatletter
\def\@citex[#1]#2{\if@filesw\immediate\write\@auxout
        {\string\citation{#2}}\fi
\def\@citea{}\@cite{\@for\@citeb:=#2\do
        {\@citea\def\@citea{,}\@ifundefined
        {b@\@citeb}{{\bf ?}\@warning
        {Citation `\@citeb' on page \thepage \space undefined}}
        {\csname b@\@citeb\endcsname}}}{#1}}
\newif\if@cghi
\def\cite{\@cghitrue\@ifnextchar [{\@tempswatrue
        \@citex}{\@tempswafalse\@citex[]}}
\def\citelow{\@cghifalse\@ifnextchar [{\@tempswatrue
        \@citex}{\@tempswafalse\@citex[]}}
\def\@cite#1#2{{\if@cghi\unskip$\null^{#1}$\else #1\fi\if@tempswa\typeout
        {warning: optional citation argument ignored: `#2'} \fi}}

\def\@biblabel#1{$\null^{#1}$}
\makeatother

\newcommand{\abs}[1]{\lvert#1\rvert}

\newcommand{\lm}[1]{\underset{#1}{\lim}}

\def\dm{\lp\begin{array}}
\def\fm{\end{array}\rp}
\def\dbb{\lb\begin{array}}
\def\fbb{\end{array}\rb}
\def\dbn{\left.\begin{array}}
\def\fbn{\end{array}\right.}

\def\lb{\left[}
\def\rb{\right]}
\def\lp{\left(}
\def\rp{\right)}

\def\m3{M_3 \lp \cc \rp}
\def\m2{M_2 \lp \cc \rp}

\def\cc{{\mathbb{C}}}
\def\rr{{\mathbb{R}}}

\def\mm{{M}}

\def\L2{L_2(\mm)}

\def\xo0{\omega^0_x}
\def\yo0{\omega^0_y}

\def\o0{\omega_0}

\def\xo0{x_\omega^0}
\def\yo0{y_\omega^0}

\twocolumn
\begin{document}
\title{\bf {\Large A brief remark on Unruh effect and causality}}
\author{Pierre Martinetti\\
{\em Instituto Superior T\'ecnico, Lisboa}\\
{\em pmartin@math.ist.utl.pt}}
\date{\small\today}
\maketitle
\begin{abstract}
Unruh effect states that the vacuum of a quantum field theory on
Minkovski space-time looks like a thermal state for an eternal
uniformly accelerated observer. Adaptation to the non eternal case
causes a serious pro\-blem: if the thermalization of the vacuum
depends on the lifetime of the observer, then in principle the
latest is able to deduce its lifetime from the measurement of the
temperature. This short note aims at underli\-ning that
time-energy uncertainty relation allows to adapt Unruh effect to
non-eternal observers without breaking causality. In particular we
show that our adaptation - the {\it diamonds's temperature}- of
Bisognano-Wichman approach to Unruh effect is causal\-ly
acceptable. This note is selfcontained but it is fully meaningful
as a complement to gr-qc/0212074 as well as a comment on
gr-qc/0306022.
\end{abstract}

\section{Introduction}

The Unruh effect\cite{unruh} states that an observer in
Minkov\-ski spacetime $M$ with constant accele\-ration $a$ and
{\it infinite lifetime} sees the vacuum of a quantum field theory
on $M$ as a thermal equilibrium state with temperature
\begin{equation}
\label{effetunruh}
 T_U = \frac{\hbar a}{2\pi k_b c}.
\end{equation}
This result can be obtained by observing that the vacu\-um for a
quantization scheme on all $M$ is not a pure state for an
alternative (but as well defined) quantization prescription on the
Rindler wedge $W$. The latest is physically relevant for $W$
 is the (whole and only) region of $M$ with whom an eternal uniformly
accelerated observer can interact, i.e. send a request and obtain
an answer. For a non-eternal observer $W$ has no particular
signification and the comparison between the two quantization
prescriptions is no longer significant. In this sense the eternity
of the observer is a strong requirement to derive Unruh's result.
Since no eternal observer exists this gives an equally strong
argument to question the validity of the effect\cite{fedotov}. Of
course this objection may be overcome by viewing $T_U$ as a limit
for asymptotic states\cite{leinaas}'\cite{matsas} but such a limit
is not always meaningful. For instance $T_U$ may be interpreted in
terms of Hawking radiation\cite{birrell} for eternal black holes
but not for Kerr black holes because\cite{wald} the Killing field
generating the horizon in Kerr spacetime has spacelike orbits near
infinity.

Several adaptations of $T_U$ have been
proposed\cite{diamonds}'\cite{padmanabhan}'\cite{schlicht} for an
observer with a finite lifetime
$\mathcal{T}$.\footnote{$\mathcal{T}$ is measured in the
observer's own referential} If the therma\-lization of the vacuum
survives in the non-eternal case then either the temperature $T$
does not depend on the lifetime and coherence with the eternal
case yields
\begin{equation}
\label{constante}
 T = T_U
\end{equation}
for all $\mathcal{T}$, or
\begin{equation}
\label{tl}
 T = T(\mathcal{T})
 \end{equation}
 with
 \begin{equation}
\underset{\mathcal{T}\rightarrow
 +\infty}{\lim}
 T(\mathcal{T}) = T_U.
\end{equation}
In this last case we face a severe problem: by measuring a
temperature $T_0$ an observer knowing (\ref{tl}) would be able to
deduce his lifetime $T^{-1}(T_0)$ and so he could predict the
instant of his death. How then can the temperature depend on the
lifetime without breaking causa\-lity ?

This note aims at underlining that time-energy uncertainty
relation prevents (\ref{tl}) from being automatically ruled out by
causal considerations. In particular we show that our
adaptation\cite{diamonds} of
Bisognano-Wichman's\cite{bisognano}'\cite{sewell} approach to
Unruh effect yields the highest lifetime-depending temperature
authorized by uncertainty relation.

\section{Time-energy uncertainty and\\
measurement of temperature}

The time-energy uncertainty relation\cite{cohen} states that the
time $\Delta t$ needed for a non-dissipative quantum system to
evolve in a significant manner is as long as the uncertainty
$\Delta E$ on the energy is small,
\begin{equation}
\label{uncertainty}
 \Delta t \Delta E \geq h.
\end{equation}
Detectors considered to measure Unruh temperature most often
consist in a quantum system $S$ coupled to the vacu\-um. To
measure a temperature $T$ the Unruh detector $S$ should evolve
from an initial (ground) state $E$ to an excited (thermal) state
$E + k_b T$. The energy gap $k_b T$ can be distinguished from the
ground energy level only if the latest is known with accuracy
$\Delta E < k_b T$. Consequently a significant evolution of $S$
(e.g. from the ground to an excited state) requires a period of
time not shorter than $\frac{h}{k_b T}$. Therefore an observer
with lifetime $\mathcal{T}$ is not able to measure a temperature
with accuracy greater than
$
\label{accuracy} \frac{h}{k_b\mathcal{T}}.
$
In particular if
\begin{equation}
\label{argument}
 T(\mathcal{T}) < \frac{h}{k_b \mathcal{T}}
\end{equation}
the Unruh observer has no time to measure $T$ precisely enough so
that to predict his lifetime. From this point of view a
$\mathcal{T}$-dependent tempe\-rature satisfying (\ref{argument})
is causally acceptable.

This is not a necessary condition. One may expect $T(\mathcal{T})$
to be obtained from correction of $T_U$ in powers of
$\mathcal{T}^{-1}$. If the accuracy in temperature measurement is
less than
\begin{equation}
\Delta T(\mathcal{T}) \doteq \abs{T(\mathcal{T}) - T_U}
\end{equation}
then the observer is not able to affirm that what he is measuring
is distinct from what he would measure if he was eternal. In other
words he doesn't know whether he may live forever or not. Hence
\begin{equation}
\label{precision}
 \Delta T(\mathcal{T}) <
\frac{h}{k_b\mathcal{T}}
\end{equation}
is another condition making a $\mathcal{T}$-dependent temperature
causally acceptable.

Mathematically one may have (\ref{precision}) with or without
(\ref{argument}) and vice versa. Physically what is meaningful is
first to check (\ref{argument}) (does the observer have enough
time to distinguish $k_bT$ from the ground energy ?) and, in case
the answer is no, check (\ref{precision}) (can the observer
distinguish $T$ from $T_U$ ?).
\newline

Before applying the procedure above to the diamonds's temperature,
let us discuss the interpretation of time-energy uncertainty
relation. Strictly speaking (\ref{uncertainty}) is valid for a
system described by a wave packet
and $(\Delta T)^{-1}$ is the frequency of oscillation of the
probability $P(b_m,t)$ of obtaining the eigenvalue $b_m$ in the
measurement of a given observable $B$ (not commuting with the
Hamiltonian). Whether or not a similar interpretation is valid for
the measurement of the energy during a transition between ground
and excited states is not clear to the author (but this is
certainly clarified in the suitable lite\-rature). Moreover
(\ref{uncertainty}) is valid for non dissipative system, which put
some constraint on the accelerating process of the Unruh observer.
Both restrictions can be overcome by the following considerations:
a quantum system $S$ coupled to the vacuum does not constitute by
itself an Unruh thermometer; one also needs a process to measure
the energy levels of $S$ (in the same manner that a column of
mercury alone is not a thermometer; it requires gradations marks
to be readable). To measure energy gap between quantum levels, one
applies some time-dependent perturbations in order to localize the
resonances of the system. A second version of the time-energy
relation\cite{cohen} indicates that a sinusoidal perturbation
acting for a time $\mathcal{T}$, cannot determine resonance with
accuracy greater than $\frac{\hbar}{\mathcal{T}}$. Hence
conditions similar to (\ref{argument}) and (\ref{precision}) with
$h$ replaced by $\hbar$,
\begin{eqnarray}
\label{argumentprim}
 T(\mathcal{T}) &<& \frac{\hbar}{k_b \mathcal{T}},\\
 \label{precisionprim}
 \Delta T(\mathcal{T}) &<& \frac{\hbar}{k_b\mathcal{T}}.
\end{eqnarray}
Since $\hbar < h$,  conditions (\ref{argumentprim}) and
(\ref{precisionprim}) are stronger than (\ref{argument}) and
(\ref{precision}).

\section{Diamonds's temperature}

Our adaptation of Unruh effect to bounded
tra\-jectories\cite{diamonds} is obtained by considering the
modular group\cite{hislop} associated to the region causally
connected to an non eternal observer. Concretely $W$ is replaced
by a diamond shape region $D\subset M$. Up to an acceptation of
KMS conditions\cite{haag} as a local definition of a thermal
state, the identification of the modular flow to the thermal flow
(the thermal time hypothesis\cite{carloconnes}) indicates that the
vacuum as seen by an observer with lifetime\footnote{Notations are
those of ref.[\citelow{diamonds}]: the observer's proper time
$\tau$ is measured from $-\tau_0$ to $\tau_0$.}
\begin{equation}
\mathcal{T}=2\tau_0
\end{equation}  is a thermal state whose temperature
 depends on both $\tau_0$ and the
observer proper time $\tau$,
\begin{equation}
T(\tau_0, \tau) = T_U \frac{\sinh a\tau_0}{\cosh a\tau_0 - \cosh
a\tau}
\end{equation}
where we take $c=1$. The local interpretation of KMS theory is
justified a posteriori by noting that for given $\tau_0$ and
acceleration $a$, the tempe\-rature is almost a constant for most
of the lifetime and takes the value observed in the middle of the
observer's life,
\begin{equation}
\label{temp}
 T(\tau_0, \tau) \backsimeq T(\tau_0, 0).
\end{equation}
(\ref{temp}) is called the {\it diamond's} temperature
\begin{equation}
\label{diamtemp} T_D (a, \tau_0) \doteq T_U \frac{\cosh a\tau_0 +
1}{\sinh a\tau_0}.
\end{equation}

With respect to condition (\ref{argumentprim}) diamonds's
temperature is causally acceptable for small accelerations
\begin{equation}
\label{smalla} \underset{a\rightarrow 0}{\lim}\, T_D (a,\tau_0) =
\frac{2\hbar}{\pi k_b\mathcal{T}}.
\end{equation}
For large accelerations the temperature no longer depends on the
lifetime,
\begin{equation}
\label{largea} \underset{a\rightarrow +\infty}{\lim} T_D (a,
\tau_0) = T_U.
\end{equation}
This is situation (\ref{constante}) which does not cause problems
with respect to causality. For intermediate acceleration
(\ref{argumentprim}) is not satisfied for large $\tau_0$ (see fig.
1) so we have to check for (\ref{precisionprim}).
\begin{figure}[ht]
\begin{center}
\mbox{\rotatebox{0}{\scalebox{0.75}{\includegraphics{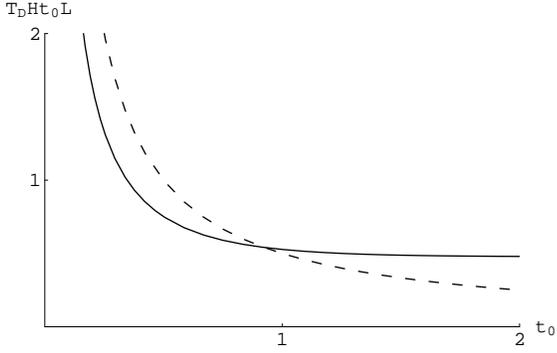}}}}
\end{center}
\caption{$T_D(\tau_0)$ for intermediate acceleration $a=3$
(vertical axe in $\frac{\hbar}{k_B}$ unit). Causally acceptable
points are under the dashed line (plot of
$\frac{1}{k_B\mathcal{T}}$).}
\end{figure}

 With
respect to conditions (\ref{precisionprim}) $T_D$ is always
accep\-table. Indeed
\begin{equation}
\Delta T(a, \tau_0) \doteq T_D(a, \tau_0) - T_U = \frac{\hbar}{\pi
k_b \mathcal{T}}
 f(a\tau_0)
\end{equation}
with
\begin{equation}
f(x)\doteq x(\frac{\cosh x + 1}{\sinh x} - 1).
\end{equation}
Since
\begin{equation}
\lm{x\rightarrow 0}\, f(x) = 2 \, , \, \lm{x\rightarrow +\infty}
f(x) = 0
\end{equation}
and $f'$ is negative on $\rr^{*+}$ then
\begin{equation}
\label{delta} 0 \leq \Delta T(a, \tau_0) \leq \frac{2\hbar}{\pi
k_b \mathcal{T}}
\end{equation}
which satisfies (\ref{precisionprim}) for any lifetime
$\mathcal{T}$.

\section{Conclusion}

For a small acceleration, diamond's temperature $T_D$ cannot be
distinguished from the ground energy of the detector, whatever the
lifetime is. For intermediate and large accelerations $T_D$ cannot
be distinguished from $T_U$. In all cases an Unruh observer is not
able to deduce information on his lifetime from the measurement of
the vacu\-um's temperature. Thus diamond's temperature is a
causally acceptable adaptation of Unruh effect to the non eternal
case. In this framework it might be interesting to re-evaluate the
intermediate result (time-dependent) of ref.[\citelow{schlicht}]
that its author estimated as non physical.

 One may find that the argument of this note - a non-eternal observer
 does not live long enough to realize that he is not eternal -  is quite paradoxical.
 Moreover corrections to Unruh temperature seems of poor interest at first sight since they are causally
 acceptable only if they are not physically detectable. Such assertions lie in the identification
 between the observer and the quantum system interacting with the vacuum. More precisely one expects
 that
 at a given instant of its lifetime the quantum system delivers the instant temperature of the vacuum.
 A more plausible possibility is to expect the measurement to occur {\it after} the end of the system.
 For instance one may think of a particle process in which Unruh temperature, including corrections, corresponds
  to a correlation between the lifetime of the particles and their production (or disintegration) rate. Such
  process have been proposed\cite{matsas}, they give a concrete signification to Unruh temperature and seem
  more closed to experimental realization  than abstract quantum system coupled to the
  vacuum.

  Even assuming that the corrections to $T_U$ are not detectable (which seems  plausible for $T_U$ itself is
  already extremely small) the good causal-behavior of $T_D$ validates the
  application
  of Unruh effect to a non-eternal
  observer. Since it is non zero even for an inertial observer, it
  suggests that the origin of the thermalization process lies
  more in the existence of an horizon than on the acce\-leration
  itself. This questions the treatment of a diamond-shape horizon in terms of (local) entropy, as this has been
  done for Rinder horizon\cite{jacobparent}.

  Finally, note that time-energy uncertainty relation
 deals more with size orders than with exact values
 (in literature one often finds (\ref{uncertainty}) with "$\gtrsim$" rather than "$\geq$"). Since
 (\ref{smalla}) and (\ref{delta}) do
  satisfy (\ref{argumentprim}) and (\ref{precisionprim}) thanks to a
  factor $\frac{2}{\pi}\backsim1$, it appears that diamond's temperature is the maximum value that one
  can canonically assign to a finite region of Minkovski spacetime.
\newline

\noindent{\bf Acknowledgment} supported by european network
HPRN-CT-1999-00118. Thanks to C. Rovelli and H.
Culetu\cite{culetu} to have suggested the importance of
time-energy relation in this framework.

\end{document}